# Monitoring the resin infusion manufacturing process under industrial environment using distributed sensors

P.WANG[1], J.MOLIMARD[1,*], S.DRAPIER[1], A.VAUTRIN[1], J.C. MINNI[2]

[1] Mechanics and Materials Processing Dep., Structures and Materials Science Division and Laboratory Claude Goux, UMR CNRS 5146

Université de Lyon

École des Mines de Saint Etienne 42023 Saint-Étienne Cedex 02, France

[2] Hexcel Corporation SAS, 38630 Les Avenières, France

* Corresponding author: molimard@emse.fr , Tel. (+33) 477 426 648, Fax. (+33) 477 420 249

**Abstract:** A novel direct approach to detect the resin flow front during the Liquid Resin Infusion process under industrial environment is proposed. To detect the resin front accurately and verify the results, which are deduced from indirect micro-thermocouples measurements, optical fiber sensors based on Fresnel reflection are utilized. It is expected that the results derived from both techniques will lead to an improvement of our understanding of the resin flow and in particular prove that micro-thermocouples can be used as sensors as routine technique under our experimental conditions. Moreover comparisons with numerical simulations are carried out and experimental and simulated mold filling times are successfully compared.

# 1. Introduction

Recently, Liquid Composite Molding (LCM) processes have been developed as interesting alternative processes to prepregs. Especially, the resin infusion-based processes (LRI/RFI: Liquid Resin Infusion/Resin Film Infusion) are now considered as cost effective processes to make lighter thick and complex composite structures [1]. They have been developed to overcome some of the limitations encountered in Resin Transfer Moulding (RTM) [2]. The resin infusion processes can be utilized in flexible conditions, for example, in low cost open molds with vacuum bags in nylon or silicone. As these processes only require low resin pressure and the tooling is less expensive than RTM rigid molds, they are particularity suitable for many small and medium size companies.

As one type of resin infusion processes, LRI process seems quite promising. In this process (see Fig.1), resin infusion is performed through a highly permeable draining fabric placed on the top of the fibers preforms. A differential pressure is created by a vacuum at the vent of the system which leads to the resin impregnation in the compressible perform, mainly in the transverse direction. The LRI process leads to final part quality improvement since resin flow and cure are distinct. However, both dimensions and fiber volume fraction of the final piece are still not well controlled mainly because of the large variation of the perform volume when vacuum and pressure are applied. Indeed, dimensions result from a transient mechanical equilibrium between flow dominated phenomena and preform compaction which is strongly non-linear. Controlling the resin flow is therefore a main issue to master the composite parts service properties. An experimental approach is developed here to monitor the main experimental data during the LRI process, such as the resin flow, in order to improve our knowledge, characterize the physical mechanisms and compare the measured results with a numerical model [4][5] predictions.

Experimental techniques dealing with flow front characterization can mainly be found in the literature related to permeability measurement issues. The determination of permeability in the preform plane has been intensively studied by measuring the one dimensional and radial flows in resin injection processes [6-8]. Compared with the in-plane flow measurement, the transverse flow detection is however considered much more difficult. The first results for transverse permeability were published in 1971 by Stedile [9], and then this detection was extended by Parnas and Han [10, 11]. After the development of in-situ sensors such as thermistor, thermocouple and optical fiber (OF) sensor [12], this technique has been widely used to study the resin front in RTM [7, 13, 14] and resin infusion processes [15, 16]. Micrometer gauge [17] or pressure sensors [18] were also employed to characterize the infusion process. More recently, ultrasound technique [19] and OFFM (optical full field method) [20, 21] have monitored successfully the transverse resin flow during the LCM processes.

In order to reduce the complexity in the measurement while decreasing as much as possible the perturbation in the real manufacturing process, during the whole LRI process micro-thermocouple and (OF) sensor have been used in this article to monitor the resin infusion process (Figure 2-b). It was shown that these two types of sensors could not only overcome the drawbacks of existing techniques used in our industrial conditions, but also characterize the resin flow position, the filling time, the curing time and the preform temperature which are key parameters controlling the part quality manufactured through LRI process.

## 2. Monitoring LRI process by (OF) sensor and micro-thermocouples

### 2.1 Principle

#### 2.1.1 Thermocouples

Micro-thermocouples of type-K have been used in our experimental study, because they are commonly used under industrial environment. This type suites well to harsh environment and presents a wide range of temperature measurement (between –75 °C and 250 °C). One thermocouple is composed of 2 wires in Chromel/Alumel with a diameter of 79 µm. This will induce very little disruption during the infusion process (Figure 2-b). Data is acquired by an acquisition unit: Agilent 34970A, a multi-acquisition system with 20 channels for measuring electric current, voltage, and resistance. The frequency acquisition meets our requirements (maximum frequency of 20 values/s), and its resolution is 0.1°C in temperature measurement. The calibration of the measurement system has been achieved in an oven during 7 isothermal heating conditions and compared with a reference platinum probe (PT100). The resolution of this calibration is 0.1°C too.

#### 2.1.2 Fresnel's (OF) sensor

To determine the resin flow front and monitor the cure advancement, the idea is to monitor the refractive index of the resin during the filling and curing stage. In fact, the reflection coefficient $R$ between two media of refractive index $n_1$ and $n_2$ can be calculated by using Fresnel's laws [22, 23]. It is expressed as:

$$R = \left( \frac{n_1 - n_2}{n_1 + n_2} \right)^2 \qquad 1$$

Moreover, the refractive index is sensitive to density variations. Indeed, the Lorentz-

Lorenz law establishes a relation between refractive index $n$ and density $\rho$ of a liquid medium (Equation 2). Here, $R_M$ is the molar refractivity and $M$ is the molar mass.

$$\frac{n^2-1}{n^2+2} = \frac{R_M}{M}\rho \qquad 2$$

As the reflected intensity at the end of an (OF) depends on the external reflection coefficient, one can monitor the resin front and degree of cure by observing the variations of the reflected intensity $I$ according to relation 3. In this relation, $A_0$ is a transmission coefficient and $B_f$ represents the amplitude of background noise, the two constants must be determined at the beginning of each experiment [22]. Fig.3 gives a schematic view of the measurement system.

$$I = A_0 R + B_f \qquad 3$$

## 2.2 Monitoring a plate resin infusion process

### 2.2.1 Experimental Set-up

Experiments were conducted with 48 plies composite plates $[0_6\ 90_6\ 90_6\ 0_6]_s$, made up of "UD fabric" reference G1157 produced by Hexcel Corp. These carbon fabrics are plain weave with 96% of weight in the warp direction and 4% of weight in the weft direction. The preform dimensions are 335 mm × 335 mm × 20 mm. For the resin, the experimental LRI tests have been performed using an epoxy resin RTM-6. Resin is preheated to 80°C in a heating chamber before infusion. A heating plate located below the semi-rigid mold heats the whole infusion system. The external pressure prescribed over the stacking is uniform and equal to the local atmospheric pressure, induced by the vacuum ensured in the sealed system. Fig.4 shows this plate infusion test carried out by LRI process, the resin inlet and outlet are indicated in this figure. The lid helps us to obtain a homogenous temperature field during the filling and curing stage: in classical tests, the filling temperature is 120°C and the curing temperature is 180°C maintained for a period of two hours.

For the purpose of detecting simultaneously the temperature variation of the preform and the refractive index of the resin, the test consists in infusing a thick plate instrumented with one (OF) sensor and six micro-thermocouples. All of the sensors positions are shown in Fig.5. Five micro-thermocouples are located across the thickness of perform. Thermocouples 3, 4 and one (OF) sensor are embedded at center of the mid-plan, with 1 cm between each sensor. Thermocouple 1 is inserted in the inlet of the system for monitoring the temperature of resin inlet.

**2.2.2 Results**

Fig.6 gives the temperature and optical signal of the sensors located at the center of the mid-ply during the whole LRI process (the temperature evolution measured by thermocouple 3 merges exactly with the one measured by thermocouple 4 in this figure). These signals can be split into 5 parts. First, the resin arrival time and the minimum temperature can be observed at zone 1, which stands for the filling stage. It can be observed that a sudden decrease in the (OF) signal and a minimum temperature occur. We infer that at the beginning of the infusion, the extremity of the (OF) is in contact with air, so its reflection coefficient rests at a high level. During the infusion stage, the resin passes over the extremity of the (OF). Due to the instantaneous decrease of the reflection coefficient, the reflected light intensity falls down. After 350 s, as the temperature increases, the refractive index and the reflected light intensity decrease. Then, in the zone where temperature increases (zone 2), the (OF) signal should continue to decrease [22]. For purely experimental reasons, this phenomenon has not been characterized completely in this zone.

Zone 3 corresponds to the curing stage. An isothermal temperature was imposed. A temperature peak can be observed at the beginning of zone 3. The zone between the beginning and the end of this peak corresponds to the increasing reflected light intensity, which relates to the density of the resin, according to Equation 2 [24]. During this stage, the liquid resin is transformed into a solid through an exothermal reaction; it involves an increase of resin density. Once the reflected light intensity no longer changes, it can be considered that the curing phase is finished and the resin is cured at 5000 s. During the cooling stage in zone 4, the resin volume is decreased, therefore its density is increased and the signal of (OF) is also increased relatively. At the end of LRI process, both temperature and reflected light intensity stabilize in zone 5.

The purpose of this study is to characterize the resin flow during the resin infusion process. Consequently, mainly the phenomena in zone 1, *i.e.* during the filling stage, are of interest.

### 2.2.3 Analysis of the filling stage

#### 2.2.3.1 Initialization of the infusion stage

The change in the temperature of resin inlet is presented in Fig.7. The temperature falls down at 10 s, which can be considered as the beginning of test. Due to the effect of resin flow, thermocouple 1 sticks on the warmer inlet pipe, then its signal gives the temperature of both resin and inlet pipe. It varies between 90 ° C and 100 ° C during the filling stage.

**2.2.3.2 Evolution of the temperature**

The five curves in Fig.8 have similar features generally. Time 0 corresponds to the beginning of the infusion process determined from the results in the Fig.7. Generally, there are temperature differences of about 10°C at the beginning of infusion stage and 5°C in the end. The temperature during the infusion stage tends to decrease when the test begins and that the resin is left free to fill in the preform. This phenomenon confirms the presence of resin that tends to cool down the preform, because the measurements of micro-thermocouple 1 show that the temperature of the resin inlet remains below 100°C (see Fig.7). The temperature signals decrease more and more when the resin front flow is getting closer to the thermocouples. The time when the minimum temperatures are obtained can be considered as the resin front arrival at the thermocouples locations. This phenomenon will be studied in the following section (paragraph 2.2.3.3) when the period around the minimum temperature is zoomed. After this minimum is reached, the temperatures of the thermocouples increase, because the resin temperature continues increasing while it flows into the preform. When the resin reaches the mould at the bottom, it is very rapidly heated by conduction (see the signals of TC 5 and TC 6). Eventually the saturated preforms reach a stabilized temperature at about 1300 s, which is related to the filling duration of the infusion stage acquired by the industrial monitoring of the experimental apparatus.

**2.2.3.3 Detection of the resin flow by (OF) sensor**

Fig.9 shows the light intensity reflected by the end of (OF) and the temperatures nearby this fiber versus time during the filling stage. The signal reflected by the fiber / air interface is perfectly stable initially. When the resin reaches the end of the (OF), the light intensity falls down due to the instantaneous change in refractive index. After 5 seconds, the (OF) signal returns to the initial level and then falls down again. It may reveal that there is some air mix in the resin flow that is to say the resin flow may not be homogeneous. When the end of the (OF) is fully surrounded by resin, its signal becomes stable and it can be considered that the resin front reaches the center of the mid-ply at 350 s. On the other hand, regarding the micro-thermocouple, we consider that once the resin front arrives in its vicinity a minimum temperature is reached. Responses of thermocouples 3 and 4 exhibit a minimum at about 320 s (315 s for thermocouple 3 and 325 s for thermocouple 4), which represents a 9% difference between the two measurement techniques. The low mismatch could be explained by the following considerations: in order to fix the position of each thermocouple, a little quantity of resin (the same resin as the one used in the infusion test) has been used to glue the head of the thermocouple; moreover, the carbon fiber has a good thermal conductivity. Changes in the thermocouples measurements should occur earlier due to this, but the difference is acceptable regarding the characteristic of infusion duration. So, according to the times corresponding to the minimum temperature, one can estimate the time required for the resin to reach the vicinity of each micro-thermocouple placed in the preform under our experimental conditions. The first prediction of the resin arrival time through the thickness of the preform by each thermocouple in the closed lid test is figured out and presented in the Table 1. In the following, micro-thermocouples will be used to estimate resin flow front during the filling stage.

### 3. Characterization of the resin flow under different experimental environments

As indicated before, the micro-thermocouples can provide general information on the resin flow in the preform during the infusion stage. Because of their low cost and easy implementation, these sensors are good candidate to monitor the LRI manufacturing process under different experimental environments.

### 3.1 In-plane characterization in closed lid infusion test

In the present studies, two infusion tests are realized under two different experimental conditions. The same reference reinforcements and resin as used in previous test are considered here. For the first experimental condition, the preform thickness is reduced and flow front advancement and resin mass are simultaneously monitored. Then, some infusion conditions will be changed in the second test. Before resin infusion, the four edges of the preform stacking should be closed normally by some adhesives to ensure that the resin infuses first in the draining fabric and then mainly across the thickness of the preform. In order to observe a longitudinal resin flow and the evolution of responses of micro-thermocouples, we let the edge close to the resin inlet (right hand side in Fig.1) to be free. All the experimental results are presented below.

#### 3.1.1 Experimental condition 1

The composite plate is made up with 24 plies $[0_6, 90_6]_S$. This resin infusion test is carried out under the standard industrial conditions (with the lid closed). Five micro-thermocouples are placed on the mid-plane (ply 12) for monitoring the resin flow on a ply plane. TC5 is at the centre of the plane, and TC1 to TC4 are glued on the edge of this ply, 5 cm away from the plate edge (see figure 10).

Fig.11 shows the evolution of temperature versus time for these 5 thermocouples on the mid-ply. Time 0 corresponds to the beginning of the infusion process when the resin enters the inlet pipe of the system. Initially, each temperature curve is rather stable, they are between 101°C and 104°C. These 5 temperature curves have similar evolution, comparable to what was described previously. As the minimum temperatures correspond to the resin front arrival, we notice that the resin reaches the mid-ply after 120 s. It can be noted that thermocouple TC5 is reached with a delay of 55 s. This delay may be due to side effects such as "race-tracking phenomena" (see Fig.12) [25, 26], which may change the progress of the fluid front and infusion pressure [27]. The times when resin flow arrives in the vicinity of the thermocouples are reported in Tab.2. At the end of the infusion stage, temperatures are stabilized again after 550 s and the filling time measured using standard industrial tools gives 600 s.

### 3.1.2 Experimental conditions 2

In order to observe the longitudinal flow, the preform will be rather thick and the micro-thermocouples should be placed on the ply that is closer to the bottom of the prefom. 48 plies composite plates $[0_6\ 90_6\ 90_6\ 0_6]s$ are used and the micro-thermocouples are placed on the 3$^{rd}$ ply, positions of which are shown in Fig.13. TC6 is at the centre of the ply, the other thermocouples (TC7-TC10) are placed 5 cm away from the plate edge.

Fig.14 shows the changes of temperature on ply 3. We can observe that there is very little difference in the temperature (about 1°C), in a given plane, between the beginning and the end of the infusion. The temperature curves for TC6, TC7 and TC9 have similar evolution as recorded in previous tests. The measured times corresponding to the minimum temperature for these 3 thermocouples are noted in Tab.3. The arrivals of the resin front around TC7 and TC9 are very close. Resin front passes over TC6, located in the center of this ply, 115 to 140 s earlier. The two other curves in this figure (for TC8 and TC10) are presenting similar profiles, which are different from the previous tests. We have observed two zones of decreasing temperature and two minimum temperatures in these two zones. As explained, in the standard test some rubber tapes are placed on the 4 edges of the preform stacking to ensure that the resin flows transversely. For the present test, we left the side near the resin entrance free. It is therefore possible to obtain a longitudinal flow at the beginning of the infusion test, which is revealed by thermocouples TC8 and TC10. The response of TC8 and TC10 has two minimum points. It can be supposed that a small quantity of resin flows through the draining fabric and enters the preform at the beginning of the filling stage. The resin film is heated after entering the preform by the system infusion, giving the first minimum point. Then, when the second minimum temperature is acquired, we consider that the transverse resin front passes around the thermocouple. This scenario is based on the difference in the longitudinal and transverse permeability. Finally, the time when the minimum temperature is achieved on each thermocouple reveal the infusion direction: from the resin inlet side towards the outlet side. The resin front is not homogeneous along the x direction, but along the y direction, it is nearly homogeneous (directions are given in Fig.13).

**3.2 In-plane and transverse detection during LRI processes**

We have already shown the monitoring of resin flow both in planes and transverse direction during LRI process in closed lid test. In order to grasp more information about the resin flow, it is interesting to characterize the resin front both in-plane and transversely in the same infusion test under other different experimental conditions. Our principle of resin front characterization by micro-thermocouples is based on measurements of local temperature evolution; therefore two tests are chosen with very different thermal conditions: the test with opened lid and in an oven.

### 3.2.1 Opened lid test

Fig.15 shows the thermocouples positions in the preform for the test carried out with the lid left open. Fig.16 gives the changes in temperature measured by thermocouples placed across the thickness of the preform. The minimum temperature on each curve is marked in Fig.16. The time corresponding to resin front arrival can be clearly observed, it is reported in Tab.4. One can verify that the resin flows almost transversely ply by ply. The curve in Fig.17 presents the position of the resin front during filling stage estimated by the results of Fig.16. Time 0 corresponds to the time when the resin front reaches the position of TC9. The filling time corresponds to the infusion duration only for the preform (we remove about 100 s, the time for the infusion of draining fabric assessed using a scale). From Fig.17, it can be deduced that the resin infuses more slowly and with more difficulty through the thickness of the preform.

Fig.18 provides information on the resin flow in the plane of ply 3. It can be noted that a large temperature gradient appears in this test. The time of resin flow arrival estimated by each thermocouple in this ply is presented in Tab.5, a non-uniform front is observed. The resin flows quickly through the center of the ply (TC5). Compared with the previous test, with the closed lid, because temperature is higher in the center it implies a lower resin viscosity, and then faster flows in that region. In-plane resin flow is estimated using thermocouples TC1-TC4 inserted in ply 3. One observes that the resin first arrives at the upper right corner and finishes to fill in the bottom left corner. This phenomenon, observed inside the preform, is the same at the beginning of infusion, when the resin impregnates the draining fabric. We also note that the infusion is much slower across the thickness (see Fig.16) than in the plane, accordingly to corresponding permeability.

**3.2.2 Resin infusion within an oven**

In this case, we want to get a resin flow during the LRI process in a more homogeneous thermal environment. The positions of micro-thermocouples are the same as shown in Fig.15.

Temperature evolution of each thermocouple is presented in Fig.19 and Fig.20, the minimum temperature is indicated on each curve. Because of the rather homogeneous thermal condition, it is difficult to find exactly the time when the minimum value appears in some thermocouples curves. For example, TC7 shows a long zone with a very small variation of temperature. The resolution of the data acquisition system is known to be 0.1°C; this zone is virtually stable, so the time corresponding to the minimum temperature has been chosen as the initial point of this stabilization zone. The times when resin front arrives around the thermocouples is then estimated, they are noted in Tab.6 and Tab.7. For the measurements across the thickness of the preform, the same phenomenon is observed, as in the previous tests: the resin infuses ply by ply through the thickness. For the measurements in a plane, the infusion order is always from the resin entrance towards the resin exit. The filling time is about 1500 s for the whole system, with 100 s to infuse the draining fabric and 1400 to infuse the perform itself.

Fig.21 describes the resin front position estimated by micro-thermocouples across the thickness of the preform during the filling stage (the determination of the initial time is the same as in the previous tests). This curve shows the same trends as those in the opened lid test (see Fig.17): for a given pressure differential, the flow front decreases.

**3.3 Discussion**

Different manufacturing conditions generate different resin flow in resin infusion processes. The important parameters of the production are compared among three plate infusion tests with 48 plies UD fabric under different experimental environments: the closed lid test, the opened lid test and the test within an oven (see Tab.8).

In the opened lid test, the low temperature of the resin inlet has disrupted the resin flow both in the draining fabric and in the preform. Because the resin cannot flow properly during the filling stage, we need much more time to infuse completely the preform, and the final composite part presents more porosity in the preform. Consequently we can observe a thicker final plate and less resin received by the preform. On the other hand, the tests with a closed lid and within an oven have similar temperature of resin inlet. Almost the same key parameters are obtained for the process, but different preform temperatures bring some slight variations of filling time. The preform used in the closed lid test is warmer than the one infused within an oven, so a shorter filling duration has been observed. In these two infusion tests, low temperature gradient in the preform and warmer resin allow us to get the final parts with a higher fiber volume fraction than in the opened lid infusion.

The approximate positions of the resin front during filling stage under three different experimental conditions are compared in Fig.22. At the beginning of resin infusion, the curves of the tests with closed lid and within an oven increase with the same trend due to the same temperature in resin inlet and hence viscosity. Then the infusion rate in the closed lid test becomes higher than the one of the test within an oven because the preform used is warmer. The resin flows more slowly in the opened lid infusion test than in the other two tests because of a lower temperature at resin inlet and a higher temperature gradient in the preform. Finally, the different thermal conditions lead to different resin flows and then very different filling times, twice as long for other thermal environment such as in the closed lid test.

A comparison can be made between experiments and simulations of the filling stage for these cases studied here. Simulations have been realized with 1462 triangle-mixed velocity-pressure elements. In our numerical model, it has been considered that the resin viscosity is constant, so we chose the experimental results of the closed lid test and within an oven for comparison. As a first approximation, Carman-Kozeny's equation has been employed to determine the average permeability tensor of the preform [4]. The boundary conditions have been selected to represent as properly as possible the industrial environment. The resin infusion has been performed by a vacuum pressure, at 1.8 mbar. At initial conditions, the thickness of the preform is 20 mm and its fiber volume fraction is 39.5%. After compaction, computations yield a thickness of 12.7 mm while measurements give 13 mm. At the end of the filling stage, numerical simulations show that the filling time of the preform is 1095 s while the experiment yields 1200 s for the closed lid test and 1400 s for the test within an oven (after removing 100 s to infuse the draining fabric). Regarding the positions of resin front during the filling stage, we obtain generally a good correlation of the results between numerical simulation and the closed lid test (see Fig.22). The filling time of the test carried out in an oven is much longer than the simulated ones due to the lower temperature of its preform (normally, the temperature of the preform is about 120°C). In the simulations, the main problems are not only the use of an isotherm numerical model, which does not take into account the variation of the resin temperature, but also to estimate the permeability of preform by using a Carman-Kozeny's equation. On the other hand, other simulation factors can not be neglected either, such as the mesh dependency, filling algorithm, macroscopical approach… To sum up, only a global numerical framework associated with proper experiments permit us to get a deep insight into the mechanisms controlling the filling stage of resin infusion process.

**4. Conclusion and future work**

Resin flow detection has been monitored using both direct and indirect approaches. As a novel direct approach, (OF) sensor based on Fresnel reflection is used to detect the resin flow front during a plate infusion process under industrial environment. It is feasible to monitor the whole LRI process, especially to detect the resin flow and the advancement of the curing stage. Micro-thermocouples can follow the temperatures inside the preform during the process. Even if it is basically an indirect approach, the results obtained prove that they can be used to get first information on the resin flow in the preform. Because the thermocouples are little intrusive and friendly-user, this article presents a complete study about the resin flow monitoring in the preform during resin infusion processes. Firstly, a two-dimensional flow is acquired in the preform, but the flow across the thickness of the preform is always slower when compared to the one in the plane of the ply. Secondly, the race-tracking phenomenon has been met when there is weak temperature gradient in the plane; on the contrary, the resin flow depends strongly on the local temperature. Finally, the first confrontation between experimental and numerical simulation results has led to the two following points: complexity of the infusion phenomenon and encouraging correlation for the filling time in the infusion stage.

In the future work, several (OF) sensors in the different measurement points in the preform will be proposed. These results can assert further the measurements of the micro-thermocouples. Moreover, the monitoring of the curing degree will be also carried out in different plies by (OF) sensors.

List of figures

Fig.16 Evolution of the temperature measured by the thermocouples placed across the thickness of the preforms in opened lid test.

Fig.17 Resin front position estimated by micro-thermocouples across the thickness of the preform during the filling stage in opened lid test (H: thickness of the zone filled and $H_0$: thickness of the preform).

Fig.18 Evolution of the temperature measured by the thermocouples placed on the ply 3 of the preform in opened lid test.

Fig.19 Temperature evolution measured by the micro-thermocouples placed on the ply 3 during the infusion test within an oven.

Fig.20 Temperature evolution measured by the micro-thermocouples placed across the thickness of the preform during the infusion test within an oven.

Fig.21 Resin front position estimated by micro-thermocouples across the thickness of the preform during the filling stage within an oven (H: thickness of the zone filled and $H_0$: thickness of the preform).

Fig.22 Comparison of the resin front positions estimated by micro-thermocouples during infusion stage under three different experimental conditions (H: thickness of the zone filled and $H_0$: thickness of the preform).

Fig.23 Comparison of the resin front positions determined between the measurements and the numerical simulation (H: thickness of the zone filled and $H_0$: thickness of the preform).

List of tables

Tab.1 Times corresponding to the minimum temperatures measured by the thermocouples placed across the thickness of the preform.

Tab.2 Times corresponding to the minimum temperatures measured by the thermocouples placed on the mean plan in experimental condition 1.

Tab.3 Times corresponding to the minimum temperatures measured by the thermocouples placed on the ply 3 in condition 2 of closed lid test.

Tab.4 Time of resin arrival over the thermocouples across the thickness of the preform used in the opened lid test.

Tab.5 Time of resin arrival over the thermocouples on the ply 3 of the preform used in the opened lid test.

Tab.6 Time of resin arrival over the thermocouples on the ply 3 of the preform used in the infusion test within an oven.

Tab.7 Time of resin arrival over the thermocouples across the thickness of the preform used in the infusion test within an oven.

Tab.8 Comparison of the key process parameters among three infusion tests under different experimental conditions.

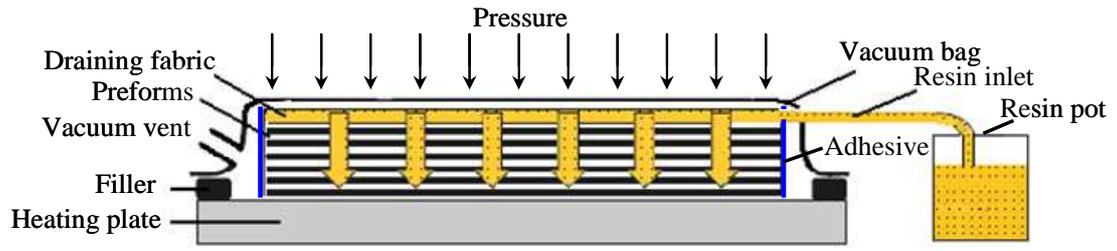

Fig.1 LRI process under standard industry conditions.

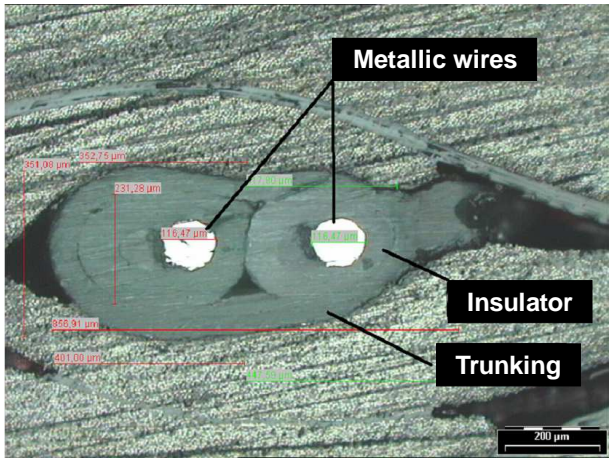 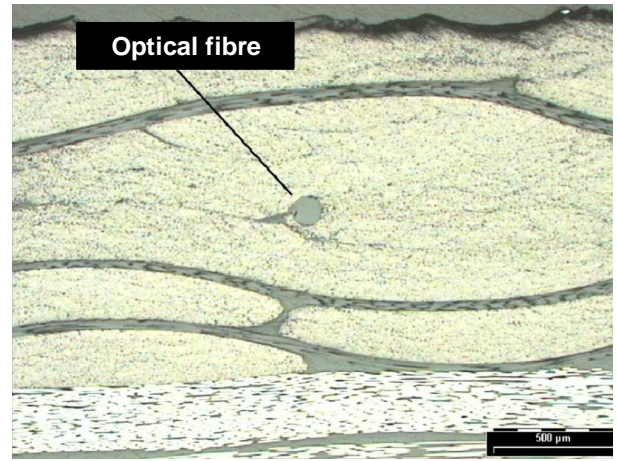

Fig.2 Embedded sensors used in this study: (a) thermocouple, (b) (OF) sensor.

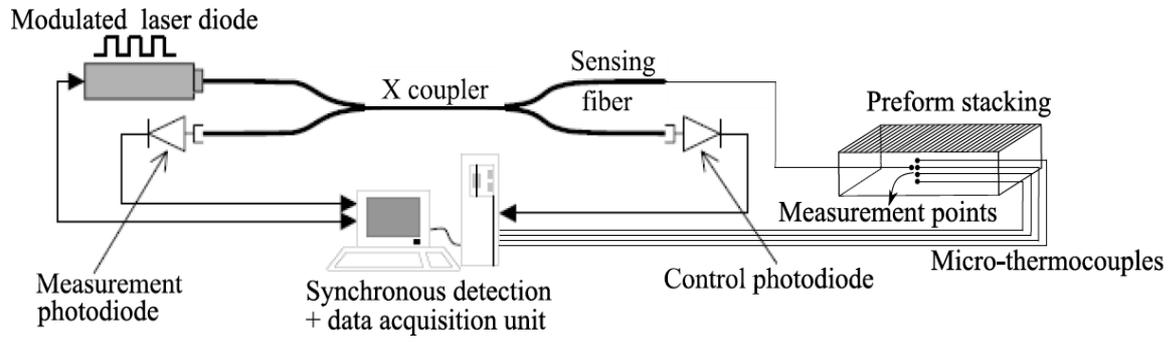

Fig.3 Principle of the measurement system by using micro-thermocouples and Fresnel's reflection with (OF) sensors.

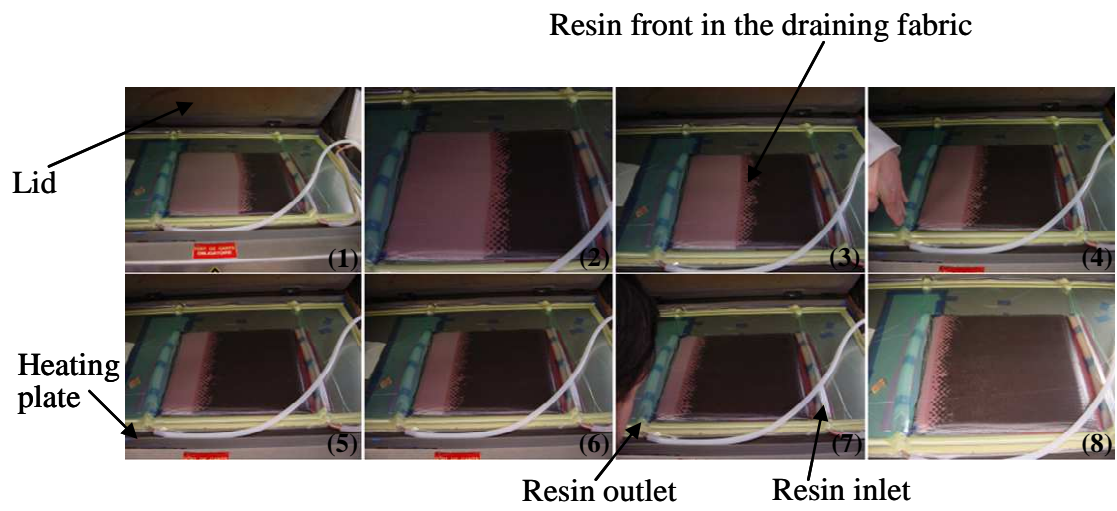

Fig.4 Infusion of a plate carried out by LRI process.

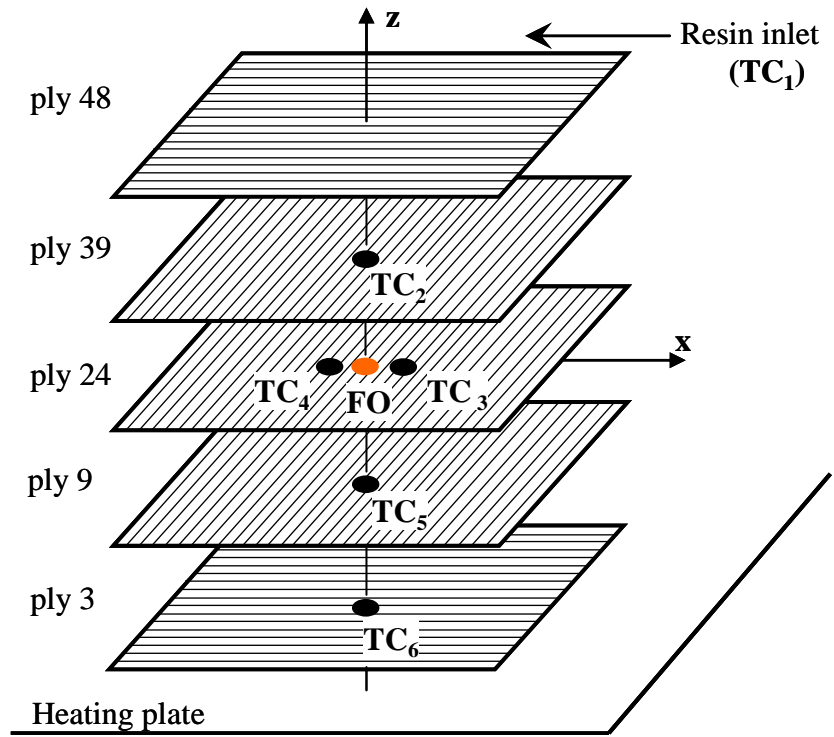

Fig.5 Positions of micro-thermocouples (TC) and (OF) sensor in the preform stacking.

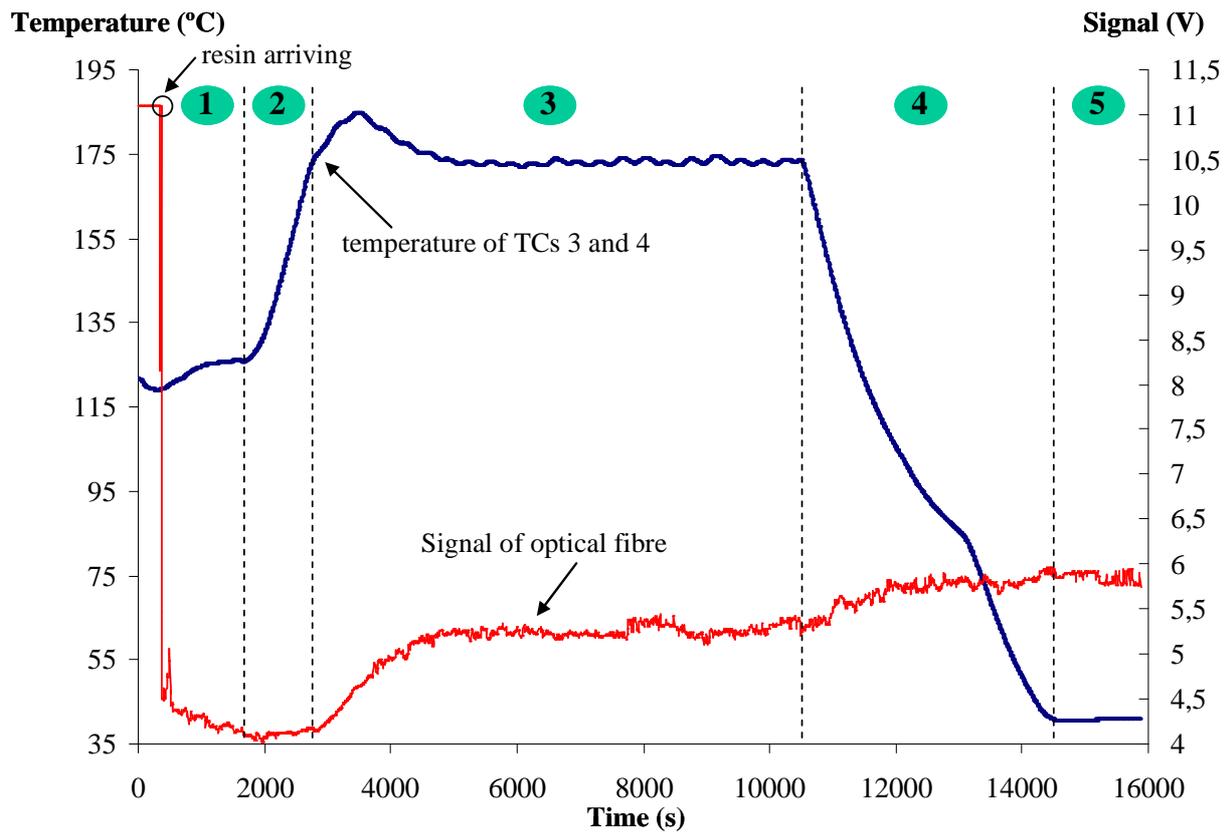

Fig.6 (OF) and the thermocouple 3 and 4 signals versus time.

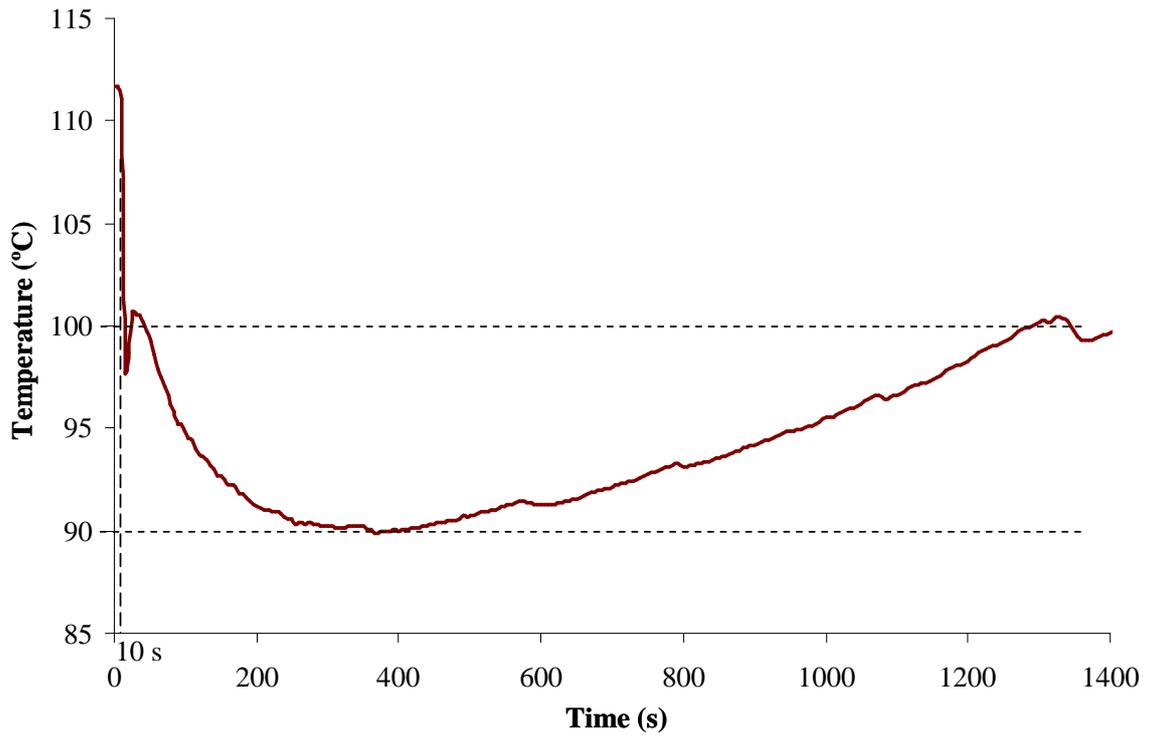

Fig.7 Change in time of the signal of the thermocouple placed in the inlet pipe of the infusion system.

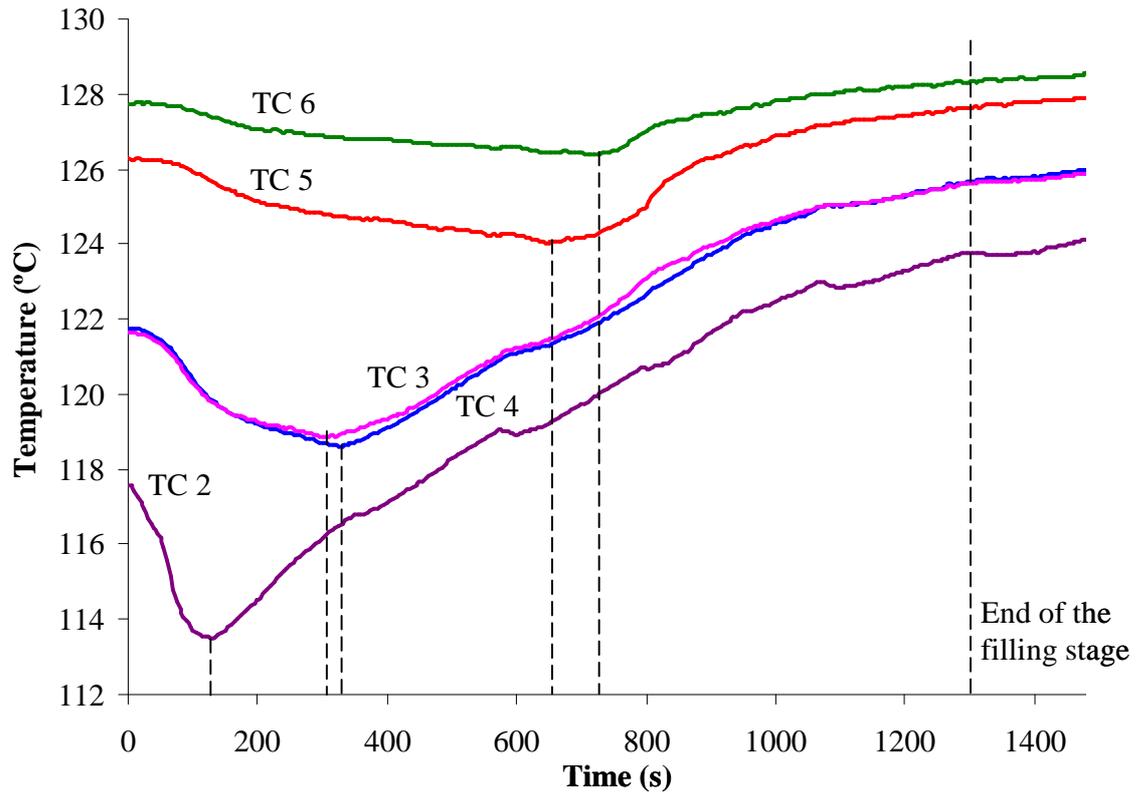

Fig.8 Change in time of the signal of the thermocouples placed across of the preform.

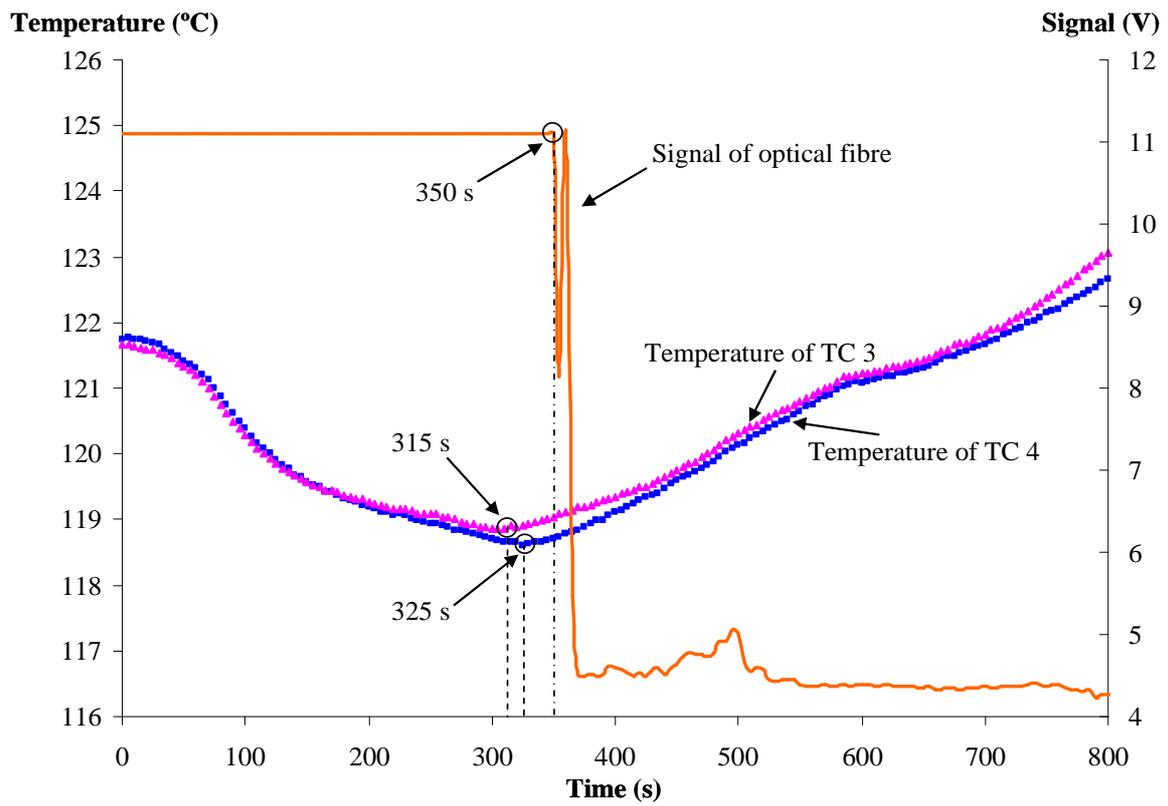

Fig.9 (OF) and the thermocouple 3 and 4 signals versus time.

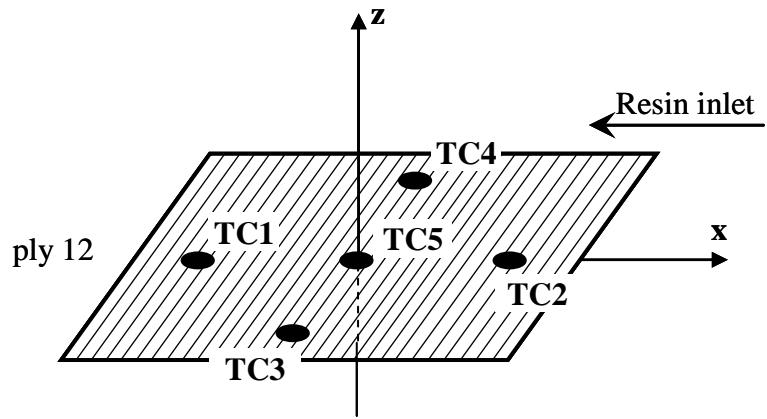

Fig.10 Positions of micro-thermocouples (TC) in the preform stacking in closed lid test of the 1$^{st}$ experimental conditions.

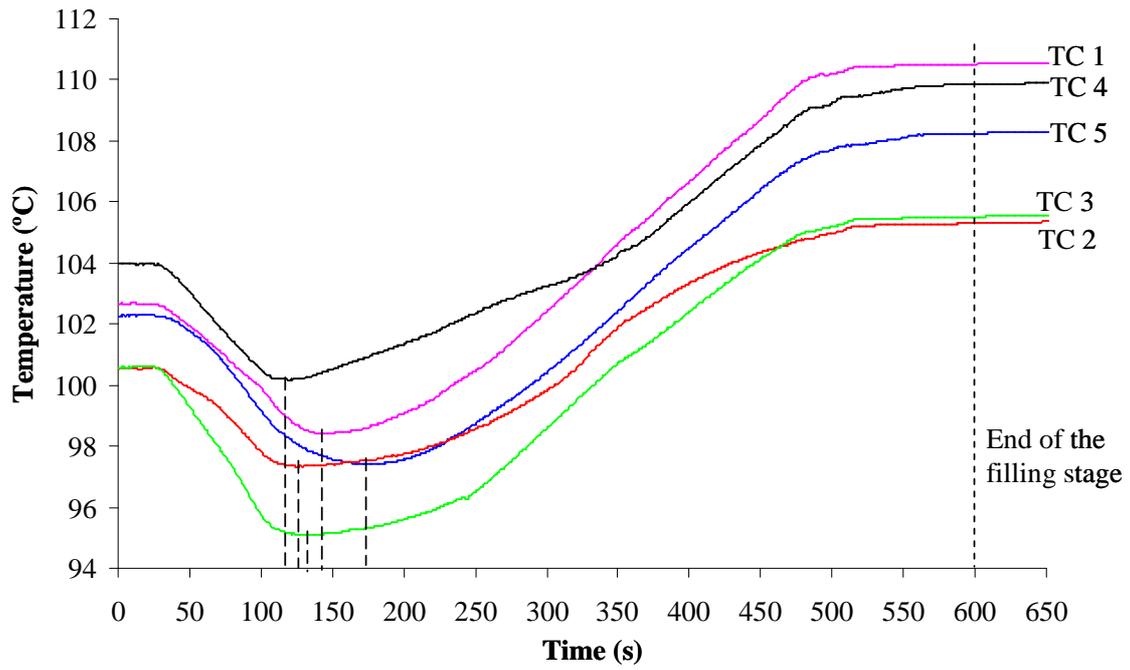

Fig.11 Evolution of the temperature measured by the thermocouples placed on the mean plan.

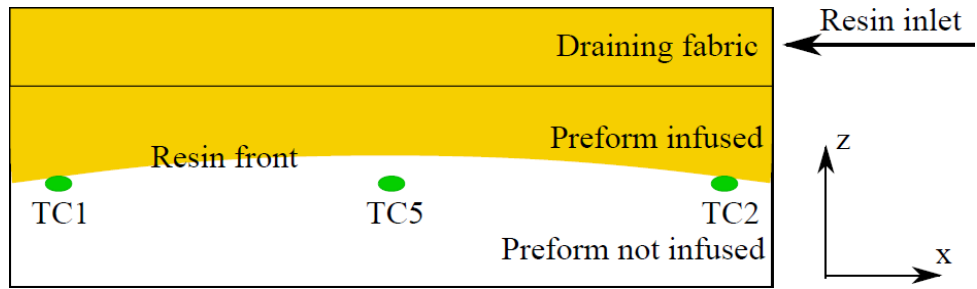

Fig.12 Resin front in the preform close to mean plan.

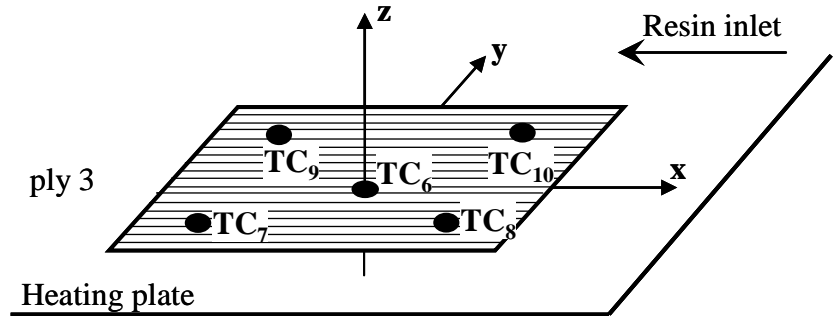

Fig.13 Positions of micro-thermocouples (TC) in the preform stacking in closed lid test of the 2$^{nd}$ experimental conditions.

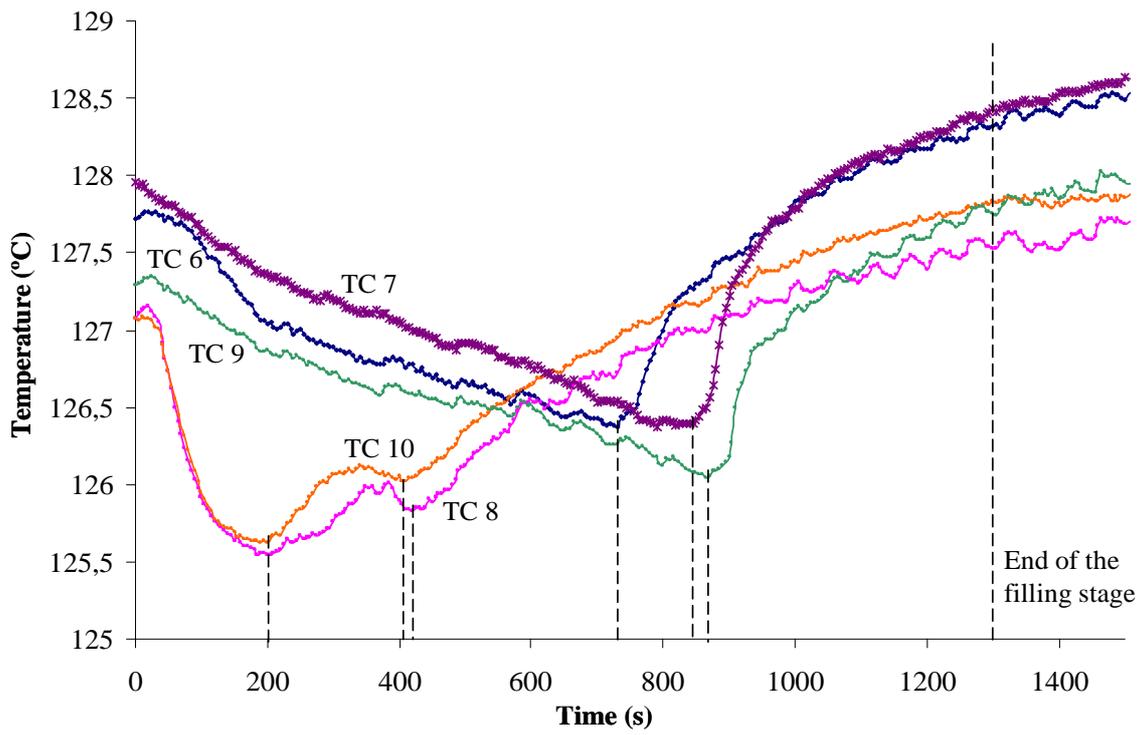

Fig.14 Evolution of the temperature measured by the thermocouples placed on the ply 3 of the preforms in closed lid test.

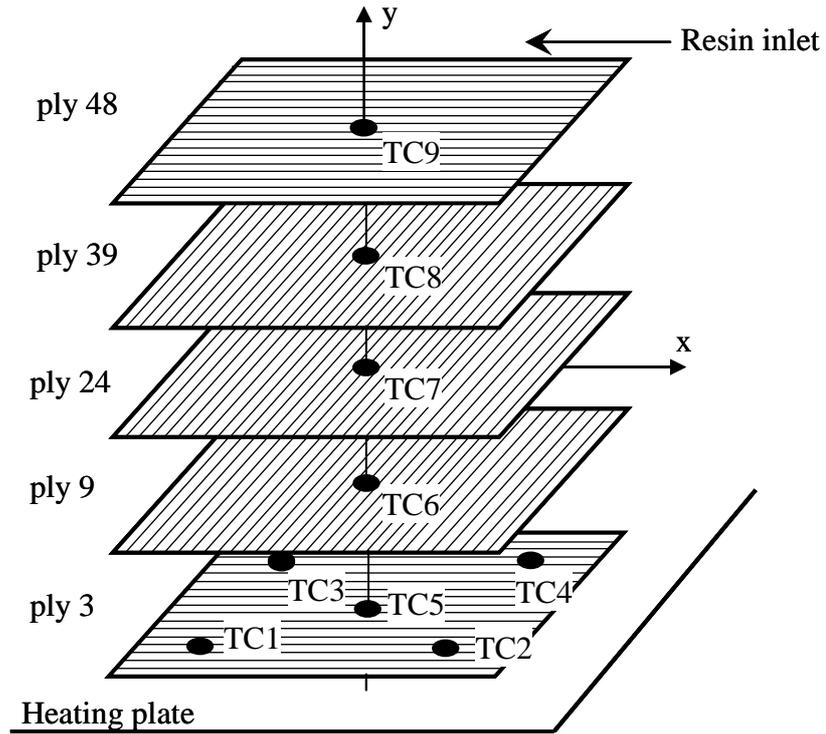

Fig.15 Positions of the micro-thermocouples (TC) embed in the preforms to detect the resin front during the filling stage in opened lid test.

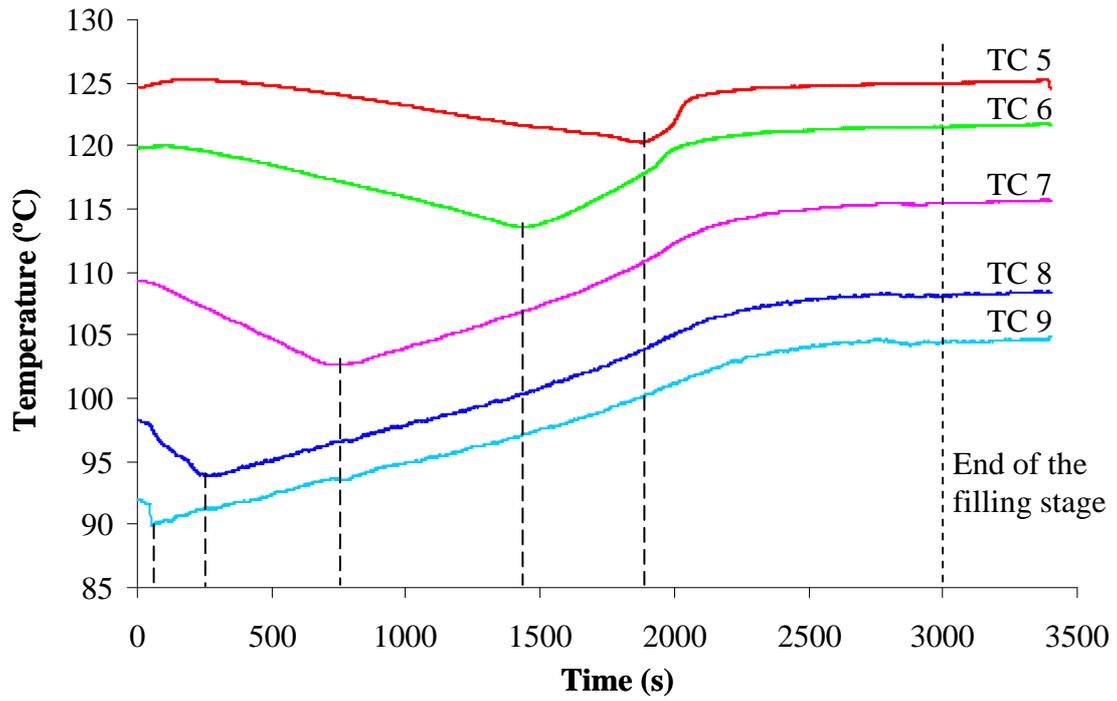

Fig.16 Evolution of the temperature measured by the thermocouples placed across the thickness of the preforms in opened lid test.

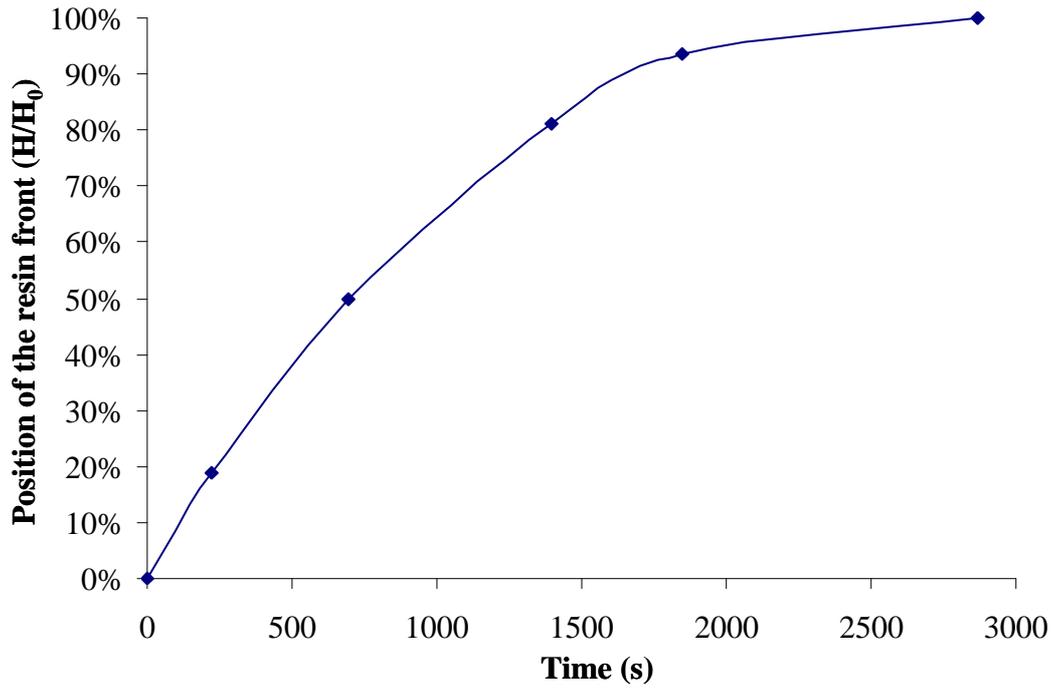

Fig.17 Resin front position estimated by micro-thermocouples across the thickness of the preform during the filling stage in opened lid test (H: thickness of the zone filled and $H_0$: thickness of the preform).

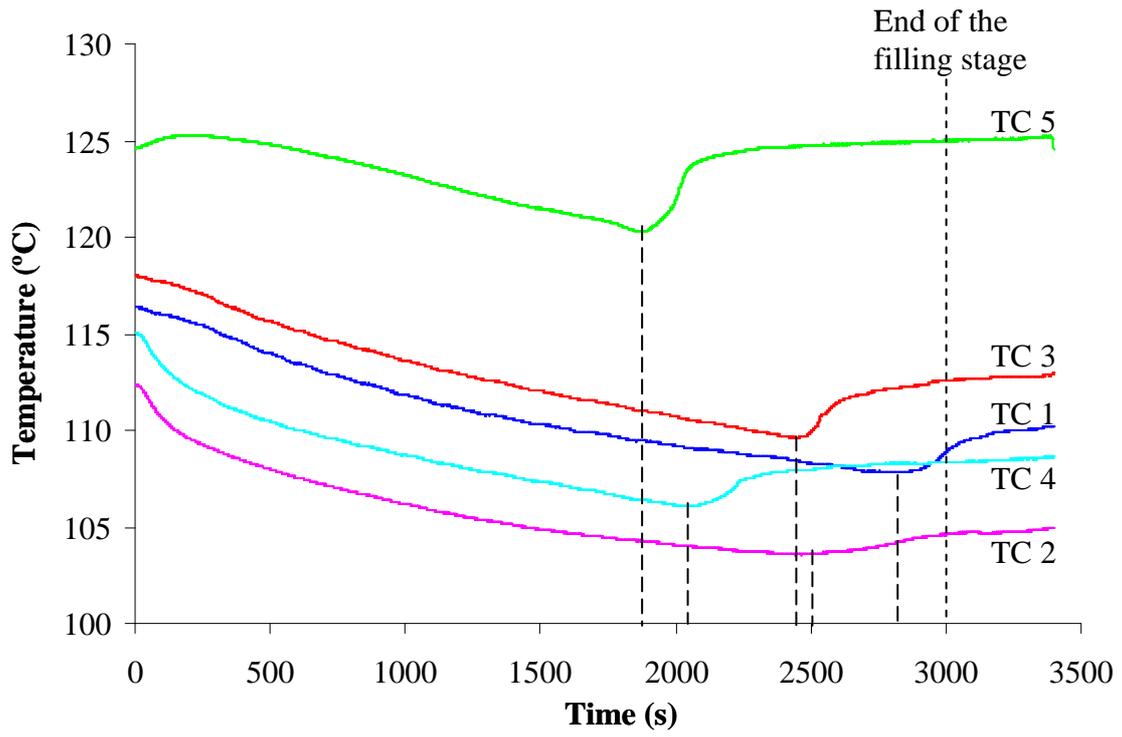

Fig.18 Evolution of the temperature measured by the thermocouples placed on the ply 3 of the preform in opened lid test.

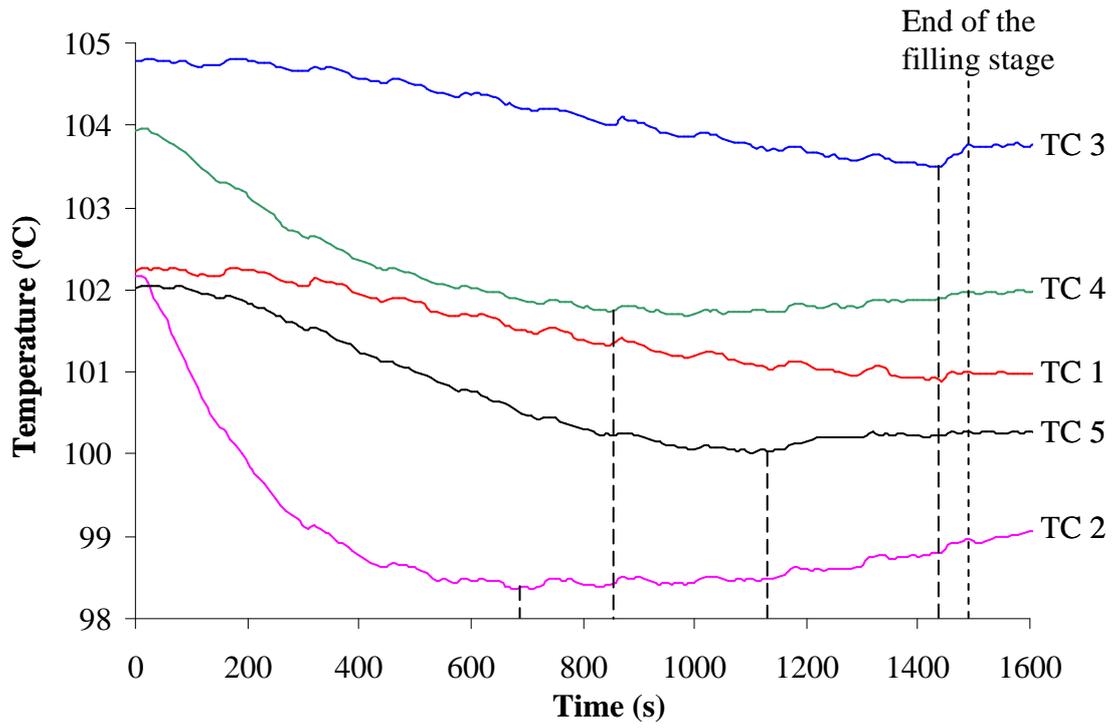

Fig.19 Temperature evolution measured by the micro-thermocouples placed on the ply 3 during the infusion test within an oven.

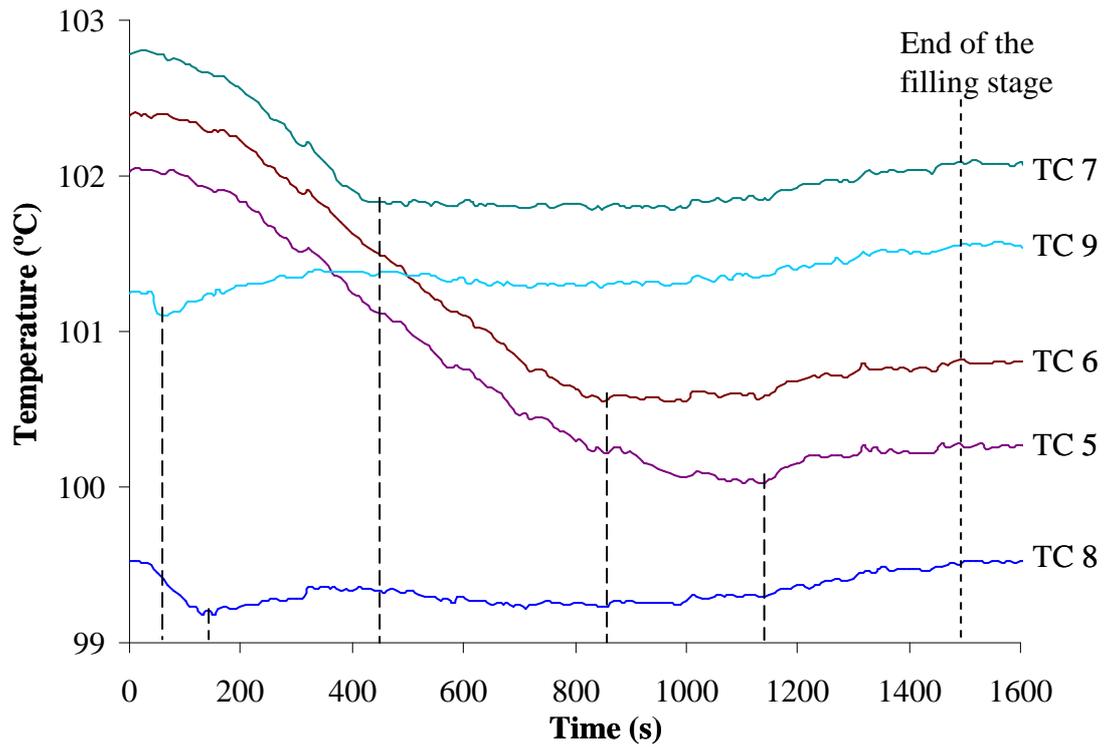

Fig.20 Temperature evolution measured by the micro-thermocouples placed across the thickness of the preform during the infusion test within an oven

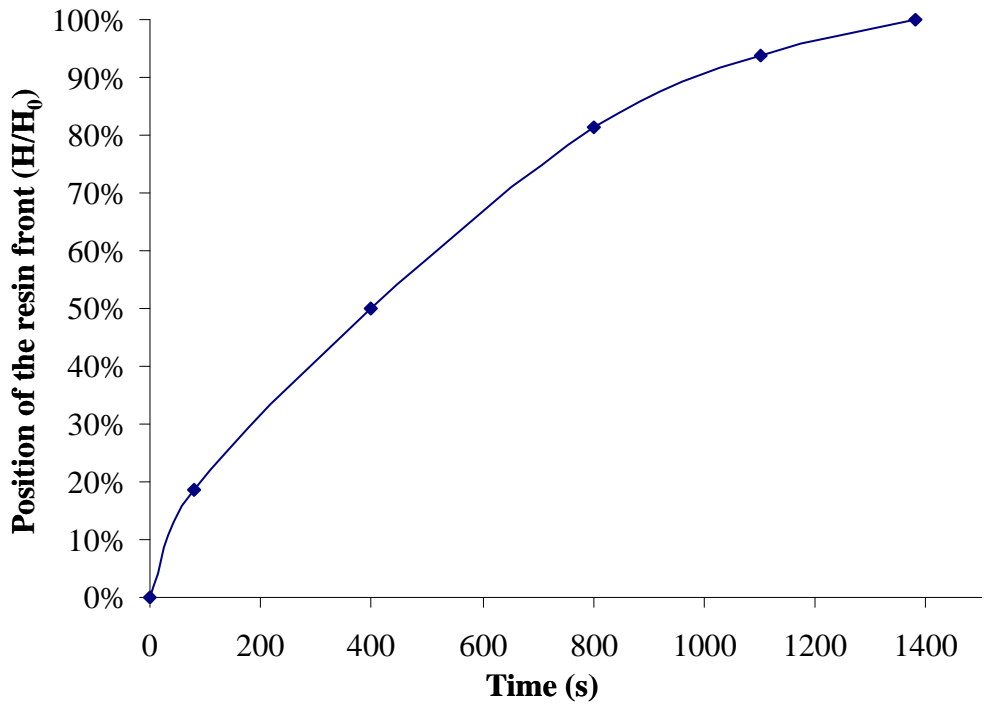

Fig.21 Resin front position estimated by micro-thermocouples across the thickness of the preform during the filling stage within an oven (H: thickness of the zone filled and $H_0$: thickness of the preform).

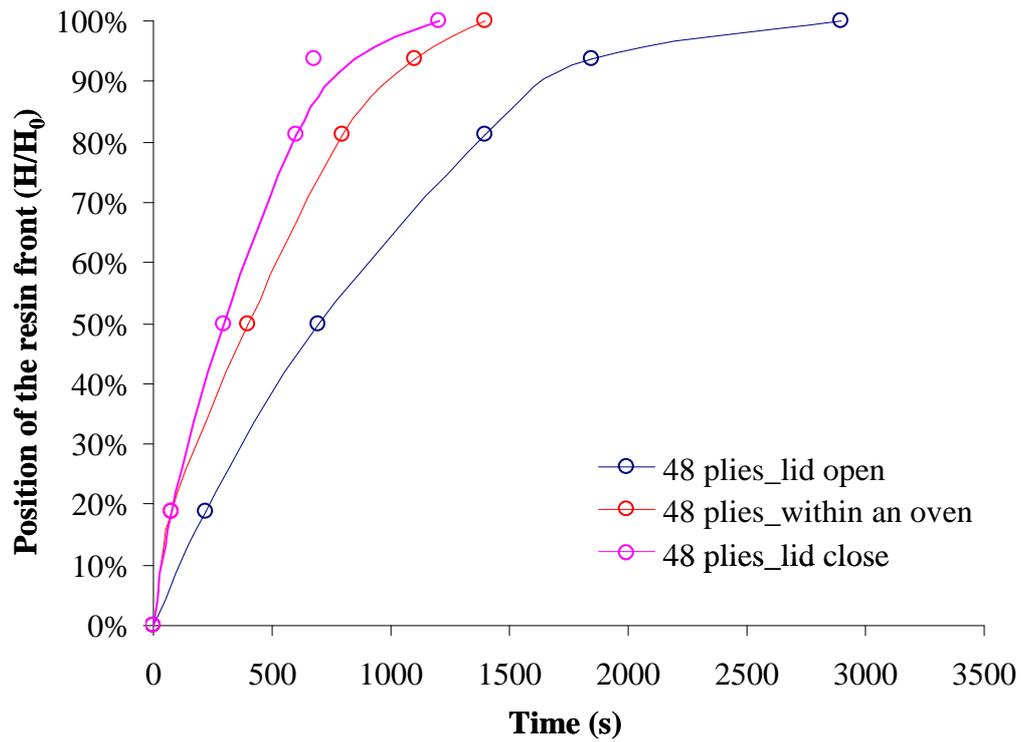

Fig.22 Comparison of the resin front positions estimated by micro-thermocouples during infusion stage under three different experimental conditions (H: thickness of the zone filled and $H_0$: thickness of the preform).

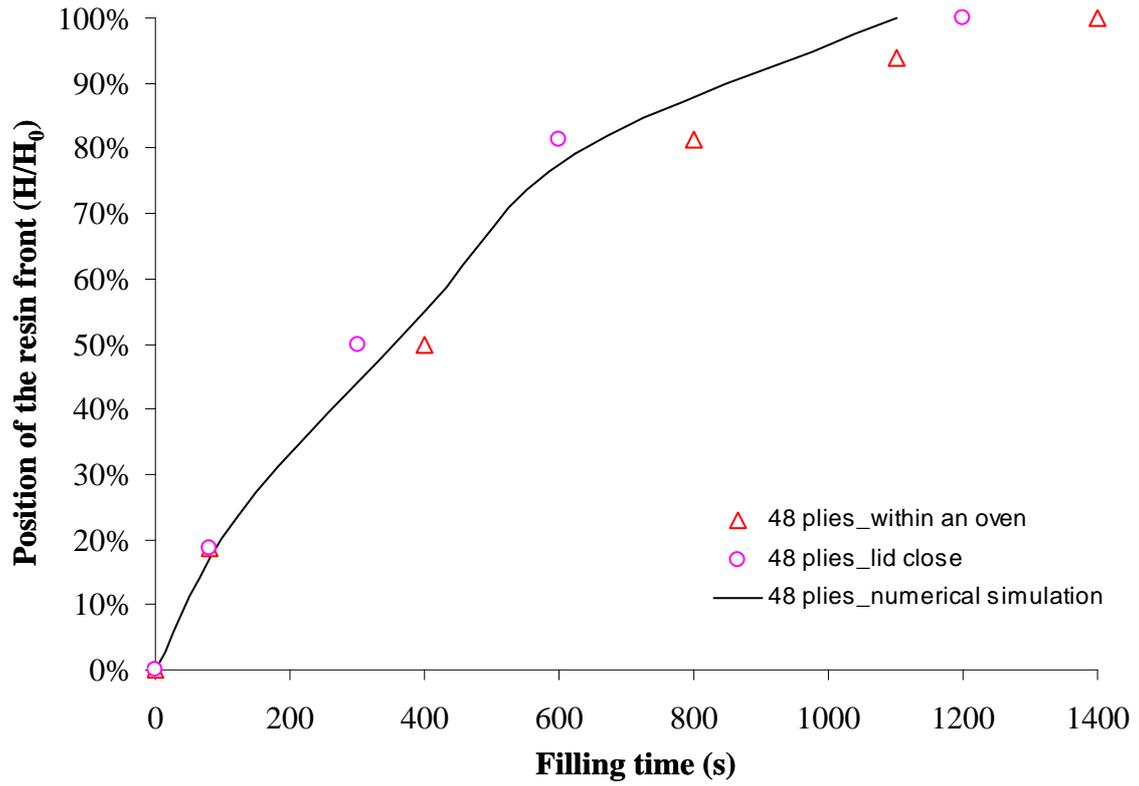

Fig.23 Comparison of the resin front positions determined between the measurements and the numerical simulation (H: thickness of the zone filled and $H_0$: thickness of the preform).

Tab.1 Times corresponding to the minimum temperatures measured by the thermocouples placed across the thickness of the preform.

| Number of thermocouples | TC2 | TC3 | TC4 | TC5 | TC6 |
|---|---|---|---|---|---|
| Time (s) | 130 | 315 | 325 | 650 | 730 |

Tab.2 Times corresponding to the minimum temperatures measured by the thermocouples placed on the mean plan in experimental condition 1.

| Number of thermocouples | TC1 | TC2 | TC3 | TC4 | TC5 |
|---|---|---|---|---|---|
| Time (s) | 145 | 125 | 130 | 120 | 175 |

Tab.3 Times corresponding to the minimum temperatures measured by the thermocouples placed on the ply 3 in condition 2 of closed lid test.

| Number of thermocouples | TC6 | TC7 | TC8 | | TC9 | TC10 | |
|---|---|---|---|---|---|---|---|
| Time (s) | 730 | 845 | 200 | 415 | 870 | 200 | 405 |

Tab.4 Time of resin arrival over the thermocouples across the thickness of the preform used in the opened lid test.

| Number of thermocouples | TC5 | TC6 | TC7 | TC8 | TC9 |
|---|---|---|---|---|---|
| Time (s) | 1900 | 1450 | 750 | 275 | 55 |

Tab.5 Time of resin arrival over the thermocouples on the ply 3 of the preform used in the opened lid test.

| Number of thermocouples | TC1 | TC2 | TC3 | TC4 | TC5 |
|---|---|---|---|---|---|
| Time (s) | 2800 | 2500 | 2450 | 2050 | 1900 |

Tab.6 Time of resin arrival over the thermocouples on the ply 3 of the preform used in the infusion test within an oven.

| Thermocouples | TC1 | TC2 | TC3 | TC4 | TC5 |
|---|---|---|---|---|---|
| Temps (s) | 1440 | 700 | 1440 | 950 | 1150 |

Tab.7 Time of resin arrival over the thermocouples across the thickness of the preform used in the infusion test within an oven.

| Thermocouples | TC5 | TC6 | TC7 | TC8 | TC9 |
|---|---|---|---|---|---|
| Temps (s) | 1150 | 850 | 450 | 130 | 50 |

Tab.8 Comparison of the key process parameters among three infusion tests under different experimental conditions.

| | Closed lid | Opened lid | Within an oven |
|---|---|---|---|
| Temperature of resin inlet | 90°C < T < 100°C | 70°C < T < 78°C | 89,5°C < T < 96°C |
| Initial temperature of the preform | ≈125°C | ≈130°C | ≈102°C |
| Filling time of the whole system (s) | 1300 | 3000 | 1500 |
| Thickness of the final composite plate (mm) | 12.02 | 12.41 | 12.08 |
| Fiber volume fraction of final plate | 62.4% | 59.8% | 62.1% |
| Resin mass in the final composite plate (g) | 550 | 520 | 540 |